\documentclass[conference]{IEEEtran}

\usepackage{amsthm}
\usepackage[cmex10]{amsmath}
\usepackage{amsfonts}
\usepackage{amssymb}

\ifCLASSINFOpdf
\usepackage[pdftex]{graphicx}
\usepackage{subfigure}
\else

\usepackage{graphicx}
\usepackage{subfigure}
\fi
\usepackage{epstopdf}
\usepackage{color}

\usepackage[font=normalsize]{caption}

\usepackage{algorithmic,algorithm}

\usepackage[numbers,sort&compress]{natbib}

\usepackage{color}

\usepackage{url}

\usepackage{mdwtab}

\begin{document}
	
	\title{A Hierarchical Approach to Encrypted Data Packet Classification in Smart Home Gateways}


	\author{\IEEEauthorblockN{Xuejiao Chen\IEEEauthorrefmark{1}, Jiahui Yu\IEEEauthorrefmark{2}, Feng Ye\IEEEauthorrefmark{3} and Pan Wang\IEEEauthorrefmark{4}}
		\IEEEauthorblockA{	\IEEEauthorrefmark{1}Department of Communication, Nanjing College of Information Technology, Nanjing, China\\	
		\IEEEauthorrefmark{2}\IEEEauthorrefmark{3}Department of Electrical \& Computer Engineering, University of Dayton, Dayton, OH, USA\\
		\IEEEauthorrefmark{4}School  of Modern Posts, Nanjing University of Posts \& Telecommunications, Nanjing, China\\			
		Emails: \IEEEauthorrefmark{1}chenxj@njcit.cn, \IEEEauthorrefmark{2}yuj016@udayton.edu, \IEEEauthorrefmark{3}fye001@udayton.edu, \IEEEauthorrefmark{4}wangpan@njupt.edu.cn}}
		
\maketitle
	
\begin{abstract}
	With the pervasive network based services in smart homes, traditional network management cannot guarantee end-user quality-of-experience (QoE) for all applications. End-user QoE must be supported by efficient network quality-of-service (QoS) measurement and efficient network resource allocation. With the software-defined network technology, the core network may be controlled more efficiently by a network service provider. However, end-to-end network QoS can hardly be improved the managing the core network only. In this paper, we propose an encrypted packet classification scheme for smart home gateways to improve end-to-end QoS measurement from the network operator side. Furthermore, other services such as statistical data collecting, billing to service providers, etc., can be provided without compromising end-user privacy nor security of a network. The proposed encrypted packet classification scheme has a two-level hierarchical structure. One is the service level, which is based on applications that have the same network QoS requirements. A faster classification scheme based on deep learning is proposed to achieve real-time classification with high accuracy. The other one is the application level, which is based on fine-grained applications. A non-real-time classifier can be applied to provide high accuracy. Evaluation is conducted on both level classifiers to demonstrate the efficiency and accuracy of the two types of classifiers. 
		
\end{abstract}

\section{Introduction}\label{sec:intro} 

Smart home has been advanced with various types of network applications~\cite{SDN_HOME2016,ITL_2018}. Network devices in a smart home can be divided into Home Automation, such as smart meters, smart lock, etc.; Health-care, such as heart rate monitor, wireless blood pressure monitor, etc.; and Entertainment, such as on-line video streaming, on-line gaming, etc.~\cite{SDN_HOME2016}. A network operator cannot provide network quality-of-service (QoS) guarantee as the end-to-end network QoS measurement cannot be performed without compromising the privacy of an end user. Even if network QoS can be measured, it is hard to guarantee user quality-of-experience (QoE) of all network services because the QoS requirements are different~\cite{HNA2013}. 

In this paper, we propose a hierarchical approach to encrypted data packet classification in smart home gateways. Successful packet classification at the gateway can improve end-to-end QoS measurement from the network operator side. Furthermore, other services such as statistical data collecting, billing to service providers, etc., can be provided without compromising end-user privacy nor security of a network. Existing work in packet classification mainly includes port detection, statistical processing, machine learning, and payload inspection~\cite{application_aware2017,SDN_DPI_TC2016,SDN_ML2017}. Port-based classifiers use the information in the TCP/IP headers of the packets to extract the port number which is associated with a particular application~\cite{application_aware2017}. However, the accuracy of this method is low because of port forwarding, protocol embedding, network address translation, etc. Payload inspection techniques are based on the analysis of information available in the payload of packets~\cite{SDN_DPI_TC2016}. It is often useless in case of encrypted packets. Machine learning and deep learning are the new approaches to classify data packet~\cite{SDN_ML2017,blackhat}. Our proposed packet classifiers are developed with deep learning, specifically, convolutional neural network (CNN) is applied to the design.

Due to high complexity of deep learning, real-time classification is a daunting task at home gateways. We propose to classify data packets using hierarchical classifiers, i.e., a service-aware classifier and multiple application-aware classifiers. A service-aware classifier is to process service layer packet, e.g., video streaming, on-line chat, on-line gaming, etc. in real-time. The results from service-level classification can be applied directly for network QoS measurement and network management. Researchers have worked on accelerating neural network computation from the hardware level. For example, the unique characteristics of non-volatile memories (NVM) was applied to accelerate the computation~\cite{8094838}. Parallelism has also been applied to speed up deep learning process~\cite{7846603}. Given fixed hardware and deep learning scheme, we propose to reduce the input size with a data interpolation scheme to further speed up the classification process. Note that the contribution of this work is the hierarchical structure of the classifiers, thus details of the data interpolation scheme are not presented. An application-aware classifier is to identify actual application of each encrypted packet, e.g., Youtube, Netflix, etc. from the service-level result. The application level classification does not require a real-time process, nonetheless, accuracy needs to be guaranteed. The application-level classification can be performed at the network operator side for other purposes, e.g., statistics collection, billing, etc.

The rest of this paper is organized as follows. Section~\ref{sec:framework} presents the proposed hierarchical approach to encrypted packet classification. Section~\ref{sec:methods} describes the CNN based packet classification scheme. Section~\ref{sec:results} presents the experimental environment and the evaluation results. Section~\ref{sec:conclusion} concludes this work.

\section{Hierarchical Approach for Encrypted Packet Classification}\label{sec:framework}

\begin{figure}[ht!]
	\centering
	\begin{tabular}{cc}
		\includegraphics[width=3.5 in]{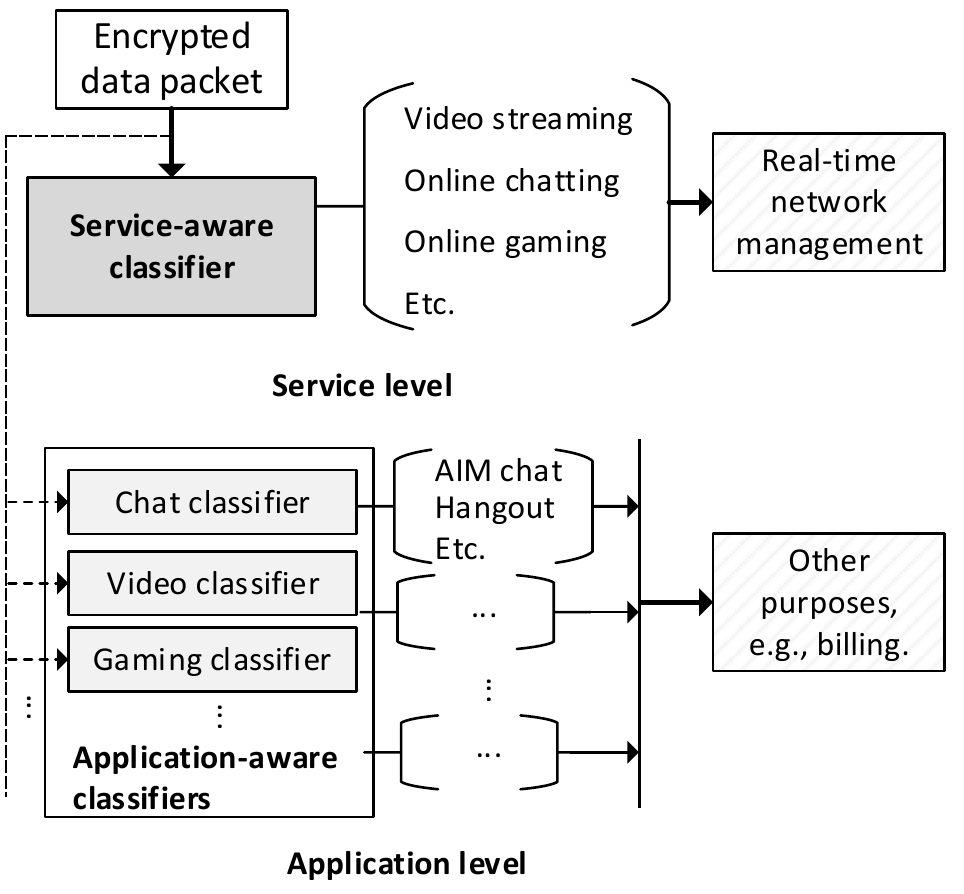}
	\end{tabular}
	\caption{Overview of the hierarchical structure.}\label{fig:framework}
\end{figure}

The proposed hierarchical structure for encrypted packet classification is shown in Fig.~\ref{fig:framework}. The structure is composed of two levels, \emph{service level} and \emph{application level}.

\subsection{Service Level}
The service level is supported by a \emph{service-aware classifier}. The classifier can conduct real-time classification based on network services, e.g., video streaming, online chatting, on-line gaming, etc. An accurate classification will enable efficient network management. Service level includes two main parts described as follows:	
	\begin{itemize}
		\item \textbf{Service Catalog:} A service catalog is to categorize different applications into multiple services. For example, Hangout and Facebook Chat would belong to the same \emph{data chat} service category. It is crucial to form and maintain a service catalog so that different application with the same network QoS requirements can be managed at this level. Based on previous research~\cite{ye_2018,ITL_2018}, we propose to form a service catalog with 3 classes, i.e., home automation, health-care and entertainment. 24 sub-classes are further defined, including video streaming, on-line chatting, on-line gaming, etc. Due to limited space, the full catalog is not provided in this work.
				
		\item \textbf{Service-Aware Classifier.} This classifier is used for encrypted packet classification according to the service catalog sub-classes. The service-aware classifier is supposed to operate in real-time. The results can be used to label each packet, e.g., in IP Header field such as DSCP of ToS, or the SDN based control message for further network measurement and resource management. We propose to develop the classifier using a deep learning approach. Details will be given in the next section.	
		
	\end{itemize}

\subsection{Application Level}

The application level is supported by a \emph{application-aware classifier}. The classifier can identify each encrypted data packet at the application level. For example, if a service has been determined as online chatting, the application-aware classifier will further classify it as AIM, Hangout, Facebook, etc. Such information could be applied to other purposes, e.g., application rating, user billing, etc. Similar to service level, application level includes two main parts described as follows:
\begin{itemize}
	\item \textbf{Application Catalog: } An application catalog is different from the service catalog. It is a fine-grained catalog. For example, Youtube, Netflix and Vimeo are maintained separately in the catalog. For simplicity, the application catalog is composed of sub-catalog, where applications belong to the same service category are in the same application level sub-catalog. 
	
	\item \textbf{Application-Aware Classifier.} This classifier is used for application level classification in each sub-catalog. Therefore, multiple classifiers will be created at the application level. For example, AIM, Hangout, etc., will be processed by the chat-only classifier, while Youtube, Netflix, etc., will be processed by the video-only classifier. In comparison to the service level, application-aware labels cannot be added to the DSCP of ToS field due to limited sizes. Therefore, we proposed to label application level results based on SDN technologies, e.g., Netflow or sFlow~\cite{6965280,7586578}. The actual labeling is beyond the scope of this work.
\end{itemize}

\begin{figure*}[h]
	\centering
	\begin{tabular}{cc}
		\includegraphics[width=6. in]{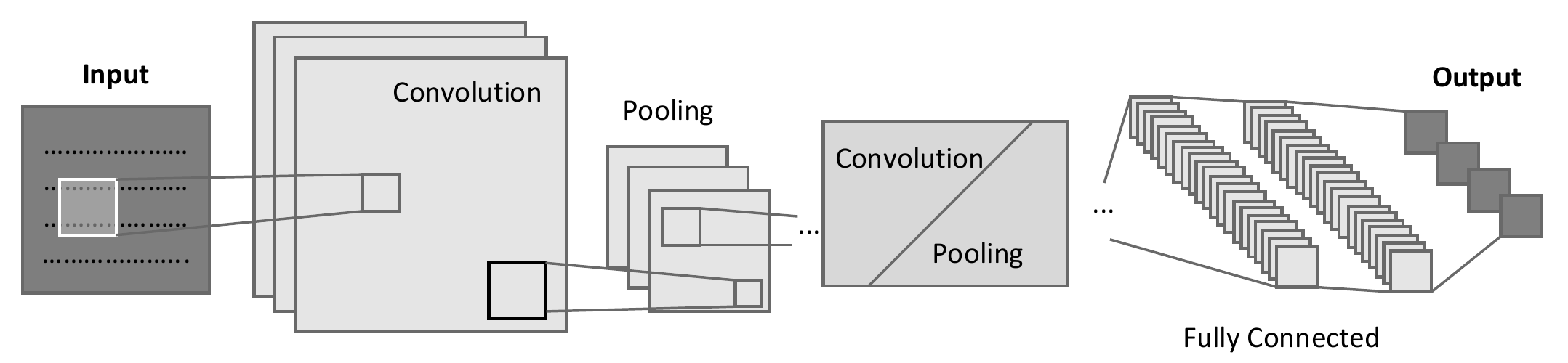}
	\end{tabular}
	\caption{The core structure of CNN used for the classifier design.}\label{fig:CNN}
\end{figure*}

\section{CNN based Encrypted Packet Classifier}\label{sec:methods}

We develop encrypted packet classifiers around CNN~\cite{Goodfellow-et-al-2016}. CNN is a typical deep learning network applied for classification. Different from a typical deep neural network, e.g., the artificial neural network, CNN applies function convolutions, as follows:
\begin{equation}
y(t) = x(t)\ast \omega(t),
\end{equation}
where $x(t)$ is the input function and $\omega(t)$ is the kernel function. 
The overview structure of a CNN is given in Fig.~\ref{fig:CNN}. As it shows, whole network consists of three types of layer: input layer, hidden layer and output layer. Computational features include local receptive field, shared weights and bias, and pooling, as described below.

\textbf{\emph{Local receptive field}}: In an ordinary neural network, the inputs are depicted as a vertical line of neurons. In CNN, the layers have neurons arranged in three dimensions: width, height, depth. Each neuron in the $l$-th layer is connected to a small region of the neurons from the previous layer. The small region is called the \emph{local receptive field}. We slide the local receptive field across the entire $(l-1)$-th layer. And there will be a different neuron in the $l$-th layer for each local receptive field.
	
\textbf{\emph{Shared weights and biases}}: transition from a layer to the next layer is defined by a weight matrix and a bias, e.g.,
\begin{equation}
f(x_i,w_i,b)=\sum_i w_i x_i+b.
\end{equation}
In CNN, the weight matrix and the bias are usually shared for different transitions. The shared weights and bias are often called as \emph{kernel} or \emph{filter}. The advantage of applying shared weights and biases is that it greatly reduces the number of parameters.
	
\textbf{\emph{Pooling}}: Pooling is an important component of a CNN. A pooling layer is usually used after a convolutional layer to reduce the dimensionality of the results from convolution. In the mean time, pooling would retain most information with the reduced dimensionality. One common procedure for pooling is \emph{max-pooling}. In max-pooling, a pooling unit outputs the largest element within a rectangular subregion. Other popular procedures for pooling include \emph{average pooling}, \emph{L2 pooling}, etc.

\begin{figure}[ht!]
	\centering
	\begin{tabular}{cc}
		\includegraphics[width=3.2 in]{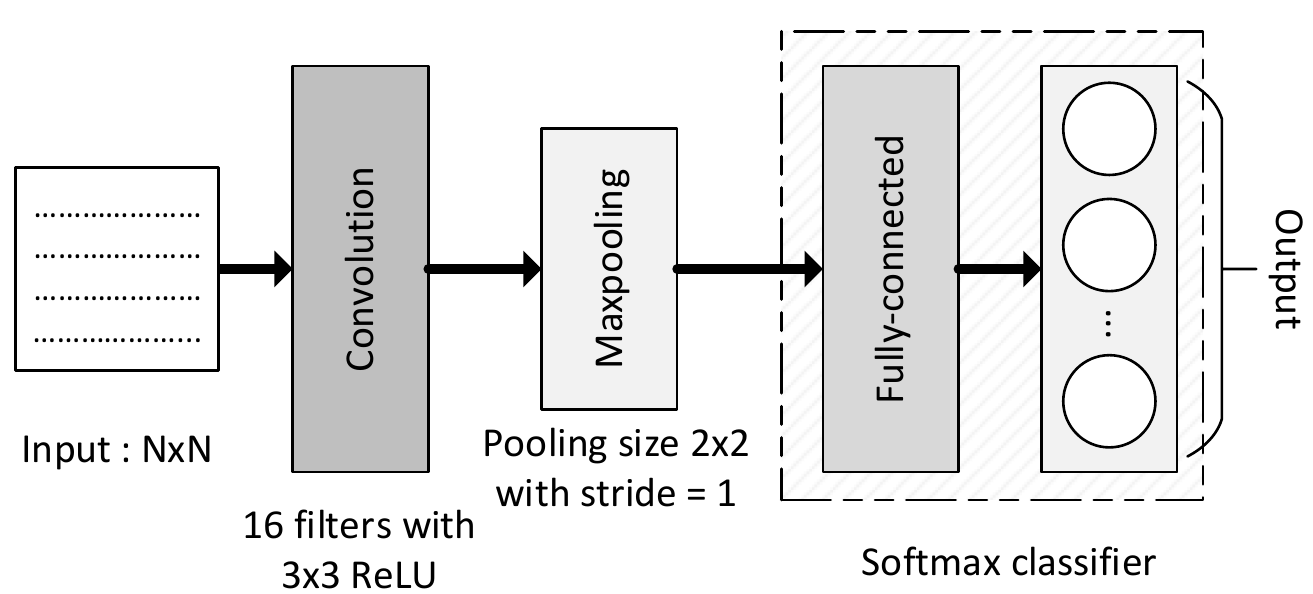}
	\end{tabular}
	\caption{Settings of the CNN based packet classifier.}\label{fig:CNN_EXP}
\end{figure}

The CNN structure specified for our proposed packet classifier is shown in Fig.~\ref{fig:CNN_EXP}. The architecture composes of one convolutional layer, one max pooling layer and a fully-connected layer with Softmax as classifier. Since the input data packet is converted into a two-dimensional (2D) matrix, we will discard depth and focus on processing the 2D data. The classification process is defined as follows:
\begin{enumerate}
	\item The convolutional layer processes the input data with 16 filters, where each filter has a size of $[3,3]$. Each filter moves 1 step after one convolution operation. 
	
	\item Results of the convolution layer are input to an activation function, i.e., a rectified linear unit (ReLU). ReLU is a non-linear operation as follows:
	Set the initial parameters: activation function $ReLU$. 
	\begin{equation}\label{eq:relu}
	ReLU(x)=\max [0,x].
	\end{equation}
	
	\item After activation of ReLU, the results are then processed through max pooling. In each step, the max pooling processes a $[2,2]$ input as follows:
	\begin{equation}\label{eq:max_pooling}
	\max \text{pooling} \left[\begin{array}{cc}
	x_1 & x_2 \\
	x_3 & x_4
	\end{array}\right]=\max(x_1,x_2,x_3,x_4).
	\end{equation}
	The max pooling has a step size of 1. 
	
	\item The Softmax classifier outputs the results as follows:
	\begin{equation}\label{eq:softmax}
	\hat{y} = \frac{\exp{z^j}}{\sum\exp{z^i}},
	\end{equation}
	where $z^j$ is the output of the $j$-th neuron. $\hat{Y} = \{\hat{y}_{1}, \hat{y}_{2}, \hat{y}_{3},.....\hat{y}_{N}\}$ is the complete set of classes, and $N$ denotes number of classes. The output with the highest probability indicates the class of the input value. 
\end{enumerate}

In the model training process, a loss function is defined based on cross entropy: 
\begin{equation}\label{eq:loss}
L = - \sum_{i=1}^{n}y_i\ln f(x_i,\theta),
\end{equation}
Stochastic gradient method is applied to find weights and bias that computes the minimum loss.

\begin{algorithm}[ht!]  
	\caption{Training the encrypted packet classifier}  \label{algorithm:CNN}
	\begin{algorithmic}[1]  
		\REQUIRE Training data, training parameters 
		\ENSURE Classification results $\hat{Y}=\{\hat{y}_{1}, \hat{y}_{2},\hat{y}_{3},\ldots\hat{y}_{N} \}$.		
		\STATE		
		\FOR {$t=1$ to $N_e$ }
		\FOR {each $M$ input data}
		\STATE For each training samples $x \in X$: 
		\STATE Compute the convolutional results;
		\STATE Compute according to Eq.~(\ref{eq:relu});
		\STATE Max pooling according to Eq.~(\ref{eq:max_pooling});
		\STATE Output classification results according to Eq.~(\ref{eq:softmax});
		\STATE Compute the training error according to Eq.~\ref{eq:loss};
		\STATE $(W,b)\leftarrow\arg\min L$;
		\ENDFOR \\ 
		\ENDFOR \\ 
	\end{algorithmic}  
\end{algorithm}

To start the training process, training parameters are set as \{$N_e$, $M$, $\eta$, $K$, $S$\}, where $N_e$ is the maximum number of Epoch, $M$ is the size of mini\_batch used in the stochastic gradient method, $\eta$ is the learning rate, $K$ is the number of filters, $S$ is the step length. The complete process for the training process is summarized in Alg.~\ref{algorithm:CNN}.

\section{Experiment and Evaluation Results}\label{sec:results}

\subsection{Experiment Settings}
In this section, we will evaluate the proposed hierarchical structure of encrypted packet classification. In particular, we will focus on the accuracy and computational efficiency in classification process. The configuration of the experiments are as follows. The evaluations are implemented using Matlab R2018a. The computer configurations are as follows: Intel i7-7700K CPU, 16GB RAM. Two GPUs are used, including an Nvidia GTX 1080 for training, and Nvidia Quatro K420 for classification tests. 

\begin{table}[ht!]
	\renewcommand\arraystretch{1.2}
	\centering
	\caption{Description about encrypted packets}\label{classification}
	\begin{tabular}{|p{0.2cm}|p{2.5cm}|p{2.5cm}|p{2.0cm}|}
		\hline
		\textbf{No} & \textbf{Application Name} & \textbf{Encryption Protocol} & \textbf{Packet Numbers} \\
		\hline
		1  & AIM\_Chat & based on HTTPS & 1243 \\
		\hline 
		2  & Facebook\_Chat & based on HTTPS & 2192\\
		\hline
		3  & Gmail\_Chat & based on HTTPS & 1800 \\
		\hline 
		4  & Hangout\_Chat & based on HTTPS & 7587\\
		\hline
		5  & ICQ\_Chat & based on HTTPS & 2721\\
		\hline  
		6  & Youtube & based on HTTPS & 12738\\
		\hline
	\end{tabular}	
\end{table}

The encrypted data is chosen from the dataset ``ISCX VPN-nonVPN traffic dataset''~\cite{ISCX}. This data set consists of several encrypted traffic data with PCAP files format. Due to limited space, data parsing process is not illustrated in this work. As a concept-proof work, we select six classes of data, as listed in Tab.~\ref{classification}. Five of the classes can be categorized as \emph{data chat}, and \emph{Youtube} represents \emph{video} service. The total amount of data is 28282 pieces. The training process is conducted with $40\%$ of the selected data set. Each data packet is 1500-byte long. The input matrix to the classier has a full size of $ 39\times 39 $. We applied data interpolation scheme to reduce the input size to $ 20 \times 20 $. Details of the data interpolation is summarized in another work of ours. For simplicity, we name a $39\times 39 $ matrix as full size data, and a $ 20 \times 20 $ as reduced size data for the rest of the evaluation.

\subsection{Evaluation of the Service-Level Classifier}

\begin{figure}[ht!]
	\centering
	\begin{tabular}{cc}
		\includegraphics[width=3.5 in]{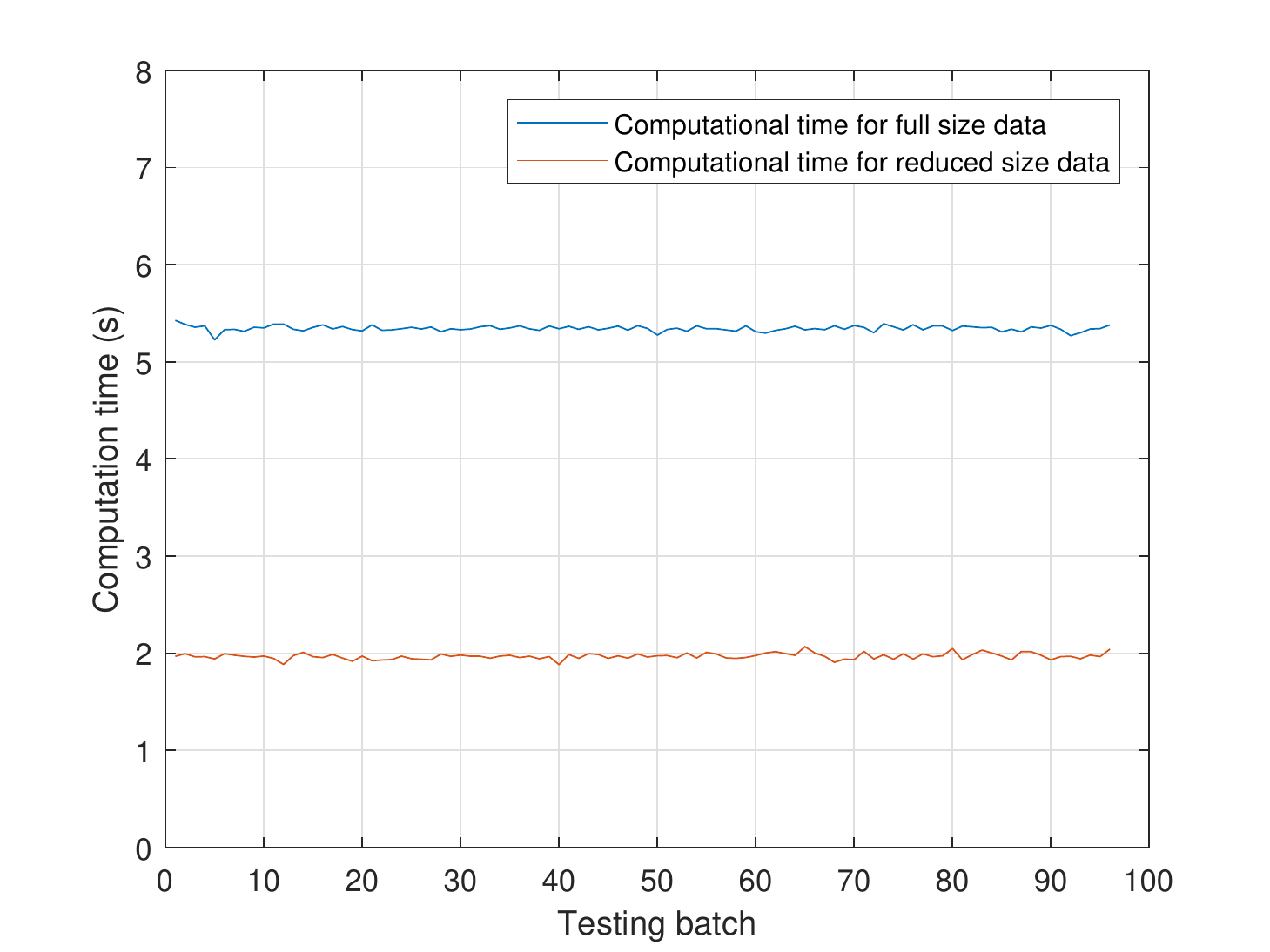}
	\end{tabular}
	\caption{Time cost of different size of matrix.}\label{fig:time}
\end{figure}

We first evaluate the service-level classifier with 2 services, i.e., data chat and video. Each time, the classifier processes $ 80\%$ i.e., $22,626$ packets). The results are extracted from 50 tests. As shown in Fig.~\ref{fig:time}, the average processing time for $22,626$ full size data is around 5.5 second, while the corresponding time to process reduced size data is around 2 seconds. The computational time is decreased by $63\%$ when processing reduced size data. Note that the classification is completed with the lower-end GPU (i.e., Quatro K420 in the setting). If processing using the higher-end GPU (i.e., GTX 1080 in the setting), the increased computational margin cannot achieve the same amount of time saving.  

We then test the accuracy of the classifiers with full and reduced size data. In each test, the same amount of data in the two service categories are randomly chosen for classification. The results are extracted from 100 tests. As shown in Tab.~\ref{accuracy2}, the average accuracy of the two services using the full-size classifier are $99.95\%$ and $99.07\%$ respectively. The average accuracy of the two services using the reduced-size classifier are $99.01\%$ and $99.25\%$ respectively. The minimum accuracy achieved in all 100 tests are $98.94\%$ and $99.15\%$ for the two service categories using the reduced-size classifier. Therefore, the proposed service-level classification can be achieved with the same accuracy and higher computational efficiency when using reduced-size classifier.

\begin{table}[ht!]
\centering
\caption{Accuracy of the 2-category classifier}
\label{accuracy2}
\begin{tabular}{|l|c|c|c|c|}
\hline  
      &\multicolumn{2}{c|}{\textbf{Chat}}&\multicolumn{2}{c|}{\textbf{Youtube}}  \\
\hline 
                & Full   &  Reduced   &  Full  &  Reduced\\
\hline 
\textbf{Maximum}  &  0.9997 & 0.9914  &  0.9916  &  0.9933 \\
\hline
\textbf{Average} & 0.9995 & 0.9901 & 0.9907 & 0.9925 \\
\hline
\textbf{Minimum} & 0.9994 & 0.9894 & 0.9897 & 0.9915 \\
\hline
\end{tabular}
\end{table}

\begin{figure}[ht!]
	\centering
	\begin{tabular}{cc}
		\includegraphics[width=3.5 in]{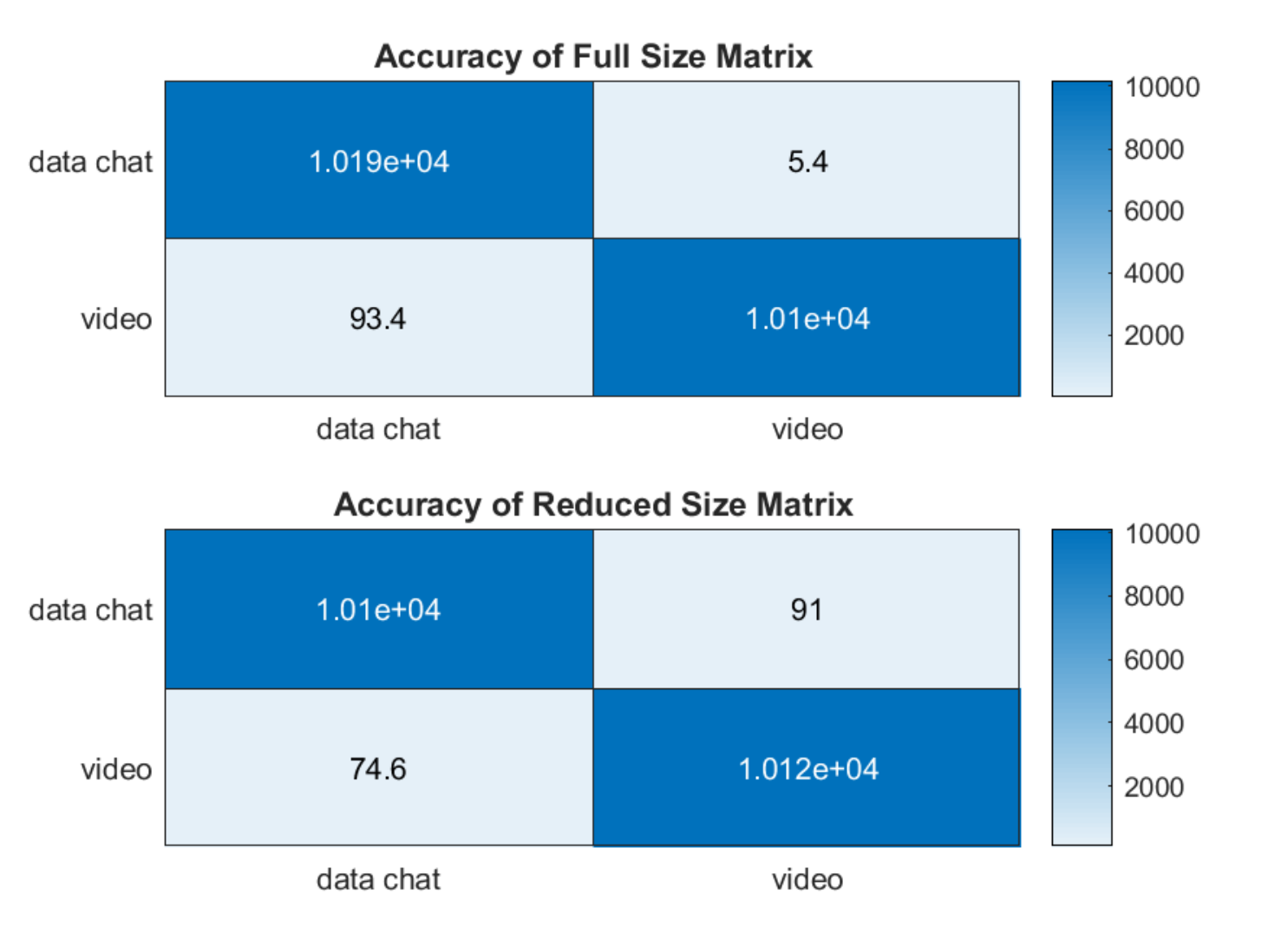}
	\end{tabular}
	\caption{Errors of classification with different matrices.}\label{fig:heat}
\end{figure}

Fig. \ref{fig:heat} shows the average of the classification results from multiple tests. When using the full size classifier, an average of $5.4$ pieces (out of $10,200$) of chat data are mis-classified into video; an average of $93.4$ pieces (out of $10,200$) of video data are mis-classified into chat class. When using the reduced size classifier, mis-classification can happen a bit more frequently, where 91 and 74.6 pieces in each category were mis-classified respectively. Nonetheless, the error rated with the reduced size classifier are only $ 0.89\% $ and $ 0.73\%$.

\subsection{Evaluation of the Application-Level Classifier}

We then demonstrate the application-level classifier. In this experiment, the classifier is to identify individual chat applications. Accuracy is the more important factor in application-aware classification. Specifically, we conduct 50 tests for testing the full-size chat-only classifier and 50 tests for the reduced-size all-application classifier. 

~\noindent
\begin{figure}[ht!]
	\centering
	\begin{tabular}{cc}
		\includegraphics[width=3.6 in]{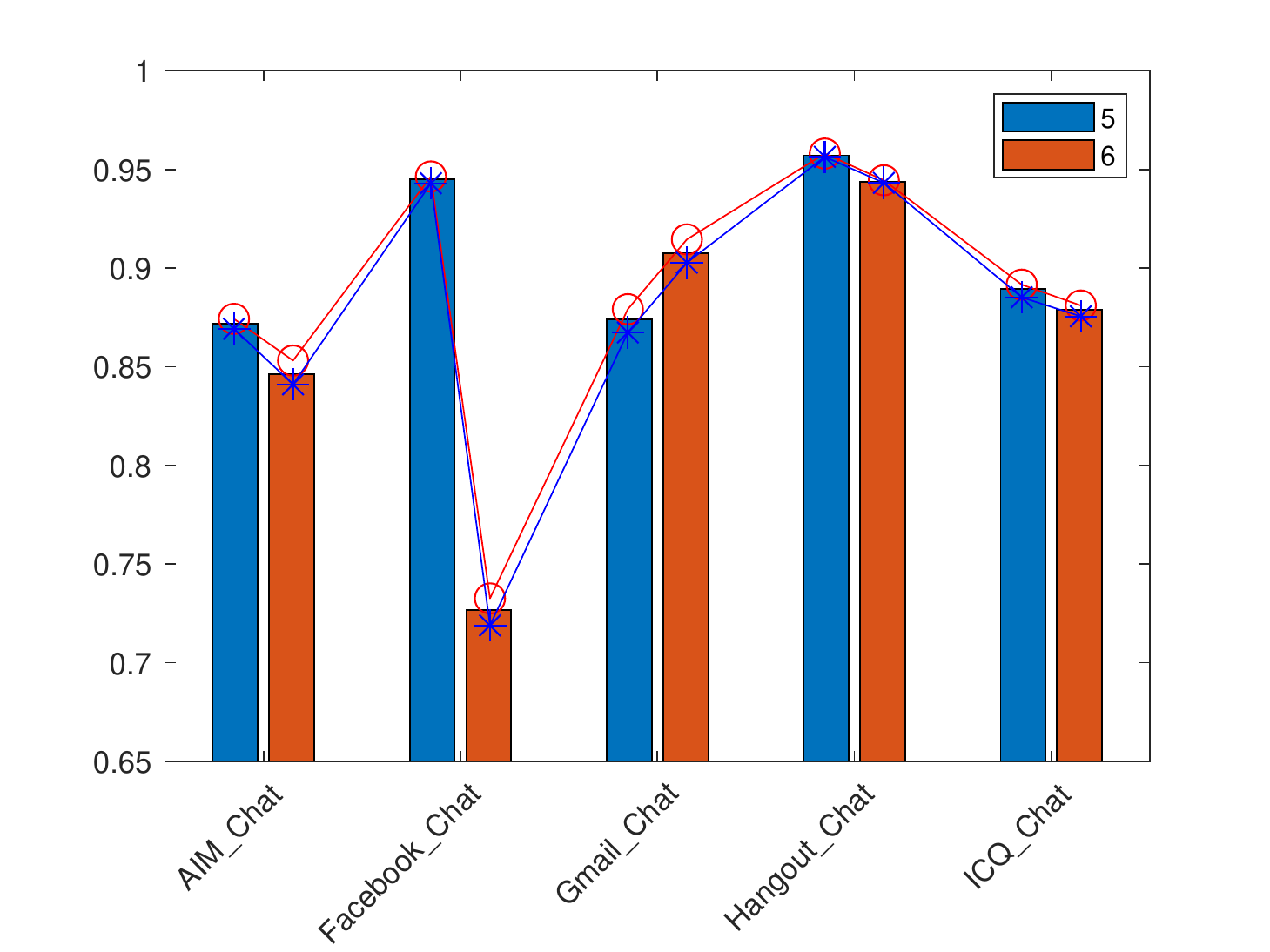}
	\end{tabular}
	\caption{Accuracy of 5 and 6 class classifier.}\label{fig:accuracy5and6}
\end{figure}

As shown in Fig.~\ref{fig:accuracy5and6}, the accuracy of the full-sized chat-only classifier (denoted as $5$ in the figure) is higher than the reduced-sized all-application classifier (denoted as $6$ in the figure) for almost every application. In particular, the accuracy is over $85\%$ for all applications when using the full-size chat-only classifier, while two applications are classified with lower accuracy when using the reduced-size all-application classifier. Therefore, in the application layer, classifiers should be trained with full-size data. Reduced-size classier should be applied only to the service layer.

\section{Conclusion}\label{sec:conclusion}

In the paper, we proposed a hierarchical approach for encrypted packet classification in smart gateways. In the proposed design, the higher level has a service-aware classifier and the lower level has multiple application-aware classifiers. At the service level, we further propose a scheme to reduce the input size, thus a gateway can compute much more efficiently for real-time classification. At the application level, application-aware classifiers can accurately identity data packets based on the actual applications. Evaluations were conducted to demonstrate the computational efficiency and accuracy at the service level. Accuracy at the application level was also demonstrated with experiments. In our future work, we will extend the work by using unsupervised deep learning approach to manage the service-level categories.

\section*{Acknowledgment}
The paper is sponsored by Jiangsu Overseas Visiting Scholar Program for University Prominent Young \& Middle-aged Teachers and Presidents, China, and Jiangsu Provincial Government Scholarship Program, China.

\renewcommand\refname{Reference}
\bibliographystyle{IEEEtran}
\bibliography{reference}

\begin{thebibliography}{10}
\providecommand{\url}[1]{#1}
\csname url@samestyle\endcsname
\providecommand{\newblock}{\relax}
\providecommand{\bibinfo}[2]{#2}
\providecommand{\BIBentrySTDinterwordspacing}{\spaceskip=0pt\relax}
\providecommand{\BIBentryALTinterwordstretchfactor}{4}
\providecommand{\BIBentryALTinterwordspacing}{\spaceskip=\fontdimen2\font plus
\BIBentryALTinterwordstretchfactor\fontdimen3\font minus
  \fontdimen4\font\relax}
\providecommand{\BIBforeignlanguage}[2]{{%
\expandafter\ifx\csname l@#1\endcsname\relax
\typeout{** WARNING: IEEEtran.bst: No hyphenation pattern has been}%
\typeout{** loaded for the language `#1'. Using the pattern for}%
\typeout{** the default language instead.}%
\else
\language=\csname l@#1\endcsname
\fi
#2}}
\providecommand{\BIBdecl}{\relax}
\BIBdecl

\bibitem{SDN_HOME2016}
P.~Gallo, K.~Kosek-Szott, S.~Szott, and I.~Tinnirello, ``Sdn@home: A method for
  controlling future wireless home networks,'' \emph{IEEE Communications
  Magazine}, vol.~54, no.~5, pp. 123--131, May 2016.

\bibitem{ITL_2018}
P.~Wang, F.~Ye, and X.~Chen, ``Smart devices information extraction in home
  wi‐fi networks,'' \emph{Internet Technology Letters}, 2018.

\bibitem{HNA2013}
A.~Kortebi, P.~L. Dain, and F.~Duré, ``Home network assistant: Towards better
  diagnostics and increased customer satisfaction,'' in \emph{Global
  Information Infrastructure Symposium - GIIS 2013}, Oct 2013, pp. 1--6.

\bibitem{application_aware2017}
S.~Jeong, D.~Lee, J.~Hyun, J.~Li, and J.~W.~K. Hong, ``Application-aware
  traffic engineering in software-defined network,'' in \emph{2017 19th
  Asia-Pacific Network Operations and Management Symposium (APNOMS)}, Sept
  2017, pp. 315--318.

\bibitem{SDN_DPI_TC2016}
G.~Li, M.~Dong, K.~Ota, J.~Wu, J.~Li, and T.~Ye, ``Deep packet inspection based
  application-aware traffic control for software defined networks,'' in
  \emph{2016 IEEE Global Communications Conference (GLOBECOM)}, Dec 2016, pp.
  1--6.

\bibitem{SDN_ML2017}
Z.~Fan and R.~Liu, ``Investigation of machine learning based network traffic
  classification,'' in \emph{2017 International Symposium on Wireless
  Communication Systems (ISWCS)}, Aug 2017, pp. 1--6.

\bibitem{blackhat}
Z.Wang, ``The application of deep learning on traffic identification,''
  \emph{Available from \url{http://www.blackhat.com}}, 2015.

\bibitem{8094838}
H.~Yan, E.~C. Ahn, and L.~Duan, ``Work-in-progress: enabling nvm-based deep
  learning acceleration using nonuniform data quantization,'' in \emph{2017
  International Conference on Compilers, Architectures and Synthesis For
  Embedded Systems (CASES)}, Oct 2017, pp. 1--2.

\bibitem{7846603}
J.~Zhang, X.~Zheng, W.~Shen, D.~Zhou, F.~Qiu, and H.~Zhang, ``A mic-based
  acceleration model of deep learning,'' in \emph{2016 International Conference
  on Audio, Language and Image Processing (ICALIP)}, July 2016, pp. 608--614.

\bibitem{ye_2018}
F.~Ye, Y.~Qian, and R.~Hu, ``Smart service-aware wireless mixed-area
  networks,'' \emph{IEEE Network Magazine}, 2018.

\bibitem{6965280}
J.~Tourrilhes, P.~Sharma, S.~Banerjee, and J.~Pettit, ``Sdn and openflow
  evolution: A standards perspective,'' \emph{Computer}, vol.~47, no.~11, pp.
  22--29, Nov 2014.

\bibitem{7586578}
L.~Huang, X.~Zhi, Q.~Gao, S.~Kausar, and S.~Zheng, ``Design and implementation
  of multicast routing system over sdn and sflow,'' in \emph{2016 8th IEEE
  International Conference on Communication Software and Networks (ICCSN)},
  June 2016, pp. 524--529.

\bibitem{Goodfellow-et-al-2016}
I.~Goodfellow, Y.~Bengio, and A.~Courville, \emph{Deep Learning}.\hskip 1em
  plus 0.5em minus 0.4em\relax MIT Press, 2016,
  \url{http://www.deeplearningbook.org}.

\bibitem{ISCX}
A.~Habibi~Lashkari, G.~Draper~Gil, M.~Mamun, and A.~Ghorbani,
  ``Characterization of encrypted and vpn traffic using time-related
  features,'' 02 2016.

\end{thebibliography}

\end{document}